\documentclass[a4paper]{jpconf}
\usepackage{graphicx}
\newcommand{\Eref}[1]{Equation (\ref{#1})}

\begin{document}
\title{XUV+IR photoionization of argon atoms: selection of sideband orders}

\author{R Della Picca$^1$, A A Gramajo$^1$, S L\'opez$^2$ and D G Arb\'o$^2$}

\address{$^1$ CONICET and Centro At\'omico Bariloche (CNEA), 8400 Bariloche, Argentina}

\address{$^2$ Institute for Astronomy and Space Physics IAFE (UBA-Conicet), Buenos Aires, Argentina}

\ead{renata@cab.cnea.gov.ar}

\begin{abstract}
We present a theoretical study of atomic ionization due to an XUV pulse in the presence
of an infrared laser. Within the strong field approximation 
and considering the periodicity and symmetry of the 
transition matrix 
we show that the photoelectron  spectrum can be described from 
the contribution during only one (or half) infrared cycle. 
These symmetry and periodicity properties impress 
selection rules which destructively cancel certain sideband orders favoring others.
In particular, we analyze the photoionization  of Argon  in four geometrical arrangements of the polarization vector and the photoelectron momentum direction.
\end{abstract}

\section{Introduction}

Laser assisted photo emission (LAPE) processes take place when typically 
extreme ultraviolet (XUV) radiation and infra red (IR) intense laser fields overlap in space and time.
Since the first theoretical prediction of "sidebands" peaks  \cite{Veniard95} a lot of experiments and theoretical studies have been performed in this area, see for example 
\cite{Itatani02, Drescher05, Maquet2007, Radcliffe2012, Meyer2012, Mazza2014,Dusterer2019}
 and references therein.
The simultaneous absorption of one high-frequency photon together with the exchange of several additional photons from the laser field lead to  equally spaced "sideband" peaks in the photo-electron (PE) spectra, located on each side of the photoionization line. 
From the theoretical point of view, the formation of these peaks can be explained as the constructive interference between electron waves emitted at different optical cycles 
of the IR laser field \cite{Kazansky10b,Gramajo18}. 
In our previous works \cite{Gramajo18, Gramajo16,Gramajo17} we have employed the semiclassical model (SCM) based on the strong field approximation (SFA) 
to identify electron trajectories and to describe the photoelectron (PE) spectrum 
as product of inter- and intracycle interferences factors, where the former accounts 
for sideband’s formation and the latter  as a modulation.

Despite the interpretation in terms of these interferences  was previously deduced employing the saddle point approximation in the temporal integration of the transition matrix, in this work we  show  that this approximation is not strictly necessary to describe the PE spectrum as the contribution of intra- and intercycle factors. 
In the special case 
that the electron emission is perpendicular to the laser polarization vector,
an alternative factorization of the PE spectrum in terms of intra-half and inter-half-cycle interferences can be obtained 
due to the initial state symmetry
\cite{Gramajo17}. 
This  results in that the  even orders of sideband peaks are canceled and the PE spectra present characteristic sideband peaks separated by twice the IR photon energy.
Experimentally, this kind of peaks suppression  was confirmed in one color above treshold ionization (ATI) of Xe at 800 nm  \cite{Korneev2012}. 
However, to the best of our knowledge there is no LAPE experiments to confirm the selection of sideband orders at particular emission directions. 
In the present  work we explore the photoionization of argon from different initial states and at different geometrical arrangements to  
theoretically establish  under what situation 
the selection of certain sideband production
can be expected.
%
High resolution experiments  under the mentioned conditions would be desirable in order to corroborate the present study. 

The paper is organized as follows: In Sec. 2, we briefly resume the SFA theory 
and analyze the properties of the temporal integral of the transition matrix.
In Sec. 3 we present the results for Argon LAPE from shell 3 and for four geometrical arrangements.
 Concluding remarks are presented in Sec. 5.
Atomic units are used throughout the paper, except when otherwise stated.

\section{Analysis and properties of the temporal integral of the transition matrix}

The theoretical approach based on the SFA is presented in detail in \cite{Gramajo18}.
 The model describes the ionization  of an atomic system (initially in the state $\varphi_i$ with ionization potential $I_p$)  
interacting with an XUV pulse assisted by a IR laser,  linearly polarized along $\hat{\varepsilon}_X$ and $\hat{\varepsilon}_L$, respectively.
The emitted electron with momentum $\vec{k}$ and energy $E=k^2/2$ is described by the Volkov wave function \cite{Volkov}.
Then, 
the photoelectron momentum distributions can be calculated as the square of the of the transition matrix $T_{if}$
\begin{equation}
\frac{dP}{d\vec{k}}=|T_{if}|^2 \qquad \textrm{where} \qquad T_{if}  =  -  \frac{i}{2}  F_{X0}  \int_{t_0}^{t_0+\tau_X}\hat{\varepsilon}_X \cdot  \vec{d} \big[ \vec{k}+\vec{A}(t)\big] \, e^{iS(t)} \,\,dt \label{Tifg} .
\end{equation}
Here the generalized action is defined as
\begin{equation}
S(t)=-\int_{t}^{\infty}dt'\left[\frac{\big( \vec{k}+\vec{A}(t')\big)^2}{2} + I_{p} -\omega_{X}\right]
\label{S1}
\end{equation}
and the dipole element is $\vec{d}(\vec{v})= \langle e^{i \vec{v}.\vec{r}} \vert \vec{r} \vert \varphi_{i}(\vec{r}) \rangle (2\pi)^{-3/2}$. 
$F_{X0}$ and $\omega_X$ are the  respective amplitude and frequency of the XUV pulse. $F_{X0}$ is nonzero only during the temporal interval $(t_0,t_0+\tau_X)$
where $\tau_X$ is the XUV pulse duration.
 During this lapse the IR 
electric field can be modeled as a cosine-like wave, hence, the
vector potential can be written as
\begin{equation}
 \vec{A}(t)=\frac{F_{L0}}{\omega_L}~\sin{(\omega_{L} t)}~\hat{\varepsilon}_L
\label{Avector}
\end{equation}
with  $F_{L0}$ the amplitude and $\omega_L$ the frequency of the laser electric field. 

In the following we analyze some properties of the transition matrix in \Eref{Tifg}.
First, we examine the action $S$ defined in \Eref{S1}, which  can be written as
\begin{equation}
S(t)= S_0 + a t + b \cos (\omega_L t) + c \sin (2\omega_L t) \label{S},
\end{equation}
where $a =  k^{2}/2+I_{p}+U_p - \omega_{X}$, $b =  -\hat{\varepsilon}_L\cdot \vec{k} F_{L0}/\omega_{L}^{2} $,  
$ c =-U_p /2\omega_{L}$, the ponderomotive energy $U_p=(F_{L0}/2 \omega_L)^2$ 
 and $S_0$ is a constant that can be omitted.
We observe that $[S(t)-a t]$ is a time-oscillating function with  the same period of the laser  $T=2\pi/\omega_L$, then
\begin{equation}
S(t+N T) = S(t)+ a N T \label{S_N}
\end{equation}
with $N$ an integer number. 
Furthermore, since the potential vector \Eref{Avector} is $T$-periodic, the dipole element is also periodic:   $\vec{d}\big[ \vec{k}+\vec{A}(t+NT)\big] =\vec{d}\big[\vec{k}+\vec{A}(t)\big] $. 

Let us introduce the quantity $I(t)$ as the contribution from zero to time $t$  as
\begin{equation}
I(t) = \int_{0}^{t}
\ell(t^\prime) \, e^{iS(t^\prime)} \, dt^\prime 
\qquad \textrm{with} \qquad \ell(t) =- \frac{i}{2} F_{X0}\hat{\varepsilon}_X \cdot\vec{d} \big[ \vec{k}+\vec{A}(t)\big]   \label{It}
\end{equation}
for $0\leq t\leq T$. 
When the integral (\ref{It}) covers all the first IR cycle, it coincides with the photoionization transition matrix for a XUV pulse that holds exactly one IR cycle
[see \Eref{Tifg}]. For this reason  we  call $|I(T)|^2 $ as the ''intra-cycle'' contribution.

\begin{figure}[]
\centering
\includegraphics[angle = 0, width=1\textwidth]{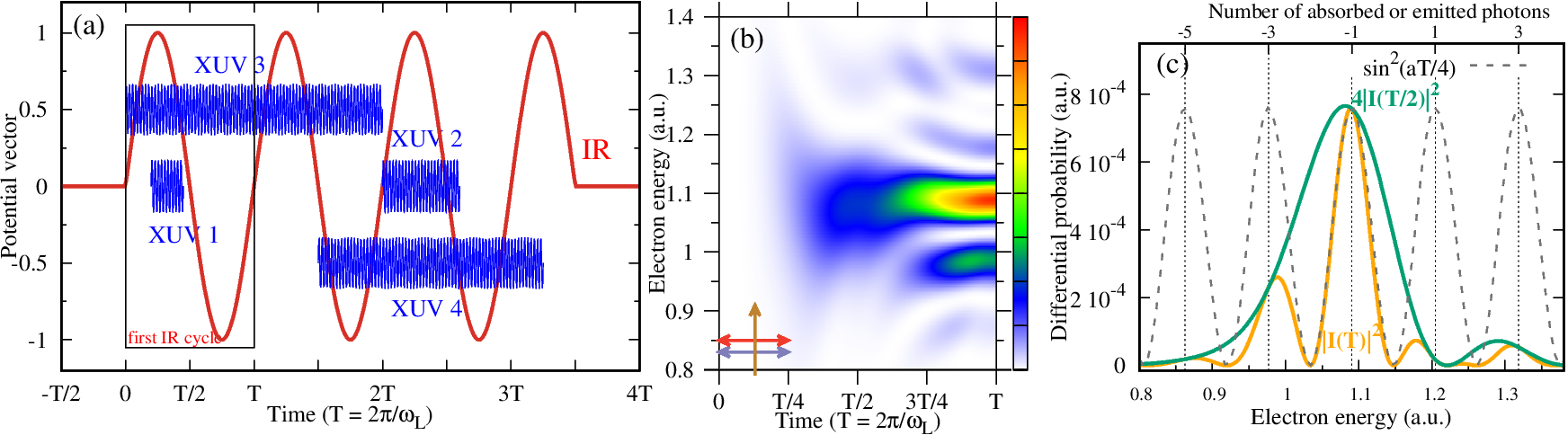}
\caption{(a) Scheme of different XUV+IR photoionization cases.
(b) Square modulus of the integral $I$  Eq. (\ref{It}) as a function of time and electron energy for the case of photoionization of Ar($3s$) with  electronic emission direction (yellow arrow) perpendicular to both polarization vectors (red -IR- and blue -XUV- horizontal arrows). (c) Cuts of plot (b) at $t=T/2$ and $T$ in arbitrary units. 
The laser parameters are  $F_{L0} = 0.041$, $\omega_L=0.057$, $F_{X0}=0.01$, $\omega_X=41 \omega_L$ and $\tau_X=3 T$.  
}
\label{fig0}
\end{figure}

We are interested in considering an arbitrary situation (i.e. arbitrary delay $t_0$ and XUV duration $\tau_X$) and to express the transition matrix  in Eq. \eref{Tifg} 
in terms of the integral $I(t)$ in \Eref{It}.
Therefore,  we analyze four cases schematized in figure \ref{fig0}(a):

\begin{description}
\item[XUV 1] (streaking): the XUV pulse is shorter than $T$  and acts during the first IR cycle.
The integration of the transition matrix
from $t_0$ to $t_0+\tau_X$ can be written as the subtraction 
of two integrals in the intervals $[0,t_0]$ and $[0, t_0+\tau_x]$:
\[
T_{if}= \int_0^{t_0+\tau_X} \ell(t) e^{iS(t)} \, dt -\int_0^{t_0} \ell(t)  e^{iS(t)} \, dt = I(t_0+\tau_X)-I(t_0).
\]
provided  that  $t_0+\tau_X\leq T$. 
\item[XUV 2] (delay of several cycles):  the XUV pulse starts at the beginning of the $N$th IR cycle ($t_0=NT$) and it is shorter than $T$, \textit{i.e},  $\tau_X \leq T$.
Performing the  transformation  $t=t^\prime+NT$ and keeping in mind the $T$-periodicity of $\ell$
and $S$ [see \Eref{S_N}] it is easy to see that 
\begin{equation}
T_{if}=\int_{NT}^{NT+\tau_X} \ell(t) e^{iS(t)} \, dt = \int_{t^\prime=0}^{t^\prime=\tau_X} \ell(t^\prime+NT) e^{iS(t^\prime+NT)}  dt^\prime  = I(\tau_X) e^{iaNT}\label{INT}.
\end{equation}
\item[XUV 3] (duration of several cycles - intercycle contribution):
 when the XUV covers several IR cycles ($\tau_X=NT$), the integral over each cycle can be summed up  using Eq. (\ref{INT}) to  obtain
\begin{eqnarray}
T_{if} &=& \int_{0}^{NT} \ell(t) e^{iS(t)} dt  = \sum_{n=0}^{N-1} \int_{nT}^{nT+T} \ell(t) e^{iS(t)} dt
       = \sum_{n=0}^{N-1} I(T) e^{ i a n T }\nonumber \\
       &=& I(T) \, \frac{\sin{(a T N /2)}}{\sin{(a T/2)}}\, e^{\left[ i a  T (N-1)/2\right] }. \label{vsduration}
\end{eqnarray}
Then, the PE momentum distribution is
\begin{eqnarray}
|T_{if}|^2&=& \underbrace{|I(T)|^{2}}_{\textrm{intracycle}} ~ \underbrace{\left[\frac{\sin{( a T N /2)}}{\sin{(a T/2)}}\right]^2}_{\textrm{intercycle}} ,  \label{intrainter}
\end{eqnarray}
where we can identify the intra- and intercycle factors. 
This kind of factorization was previously obtained within the semi classical model (SCM) in LAPE  \cite{Gramajo18} and the simple man's model in ATI \cite{Arbo2010}. 
In these works  each  contribution was recognized as the interference stemming
from trajectories within the same optical cycle (intracycle interference) and  from trajectories released at different cycles (intercycle interfernce). 
In the present work, we show in a more general way that these interferences are a consequence of the periodicity  of the transition matrix.

The zeros of the denominator in the intercycle factor 
in \Eref{intrainter}, \textit{ i.e.}, the energy values satisfying  $a T/2 =n\pi$, are  preventable singularities since the numerator also cancels, however maxima are reached at theses points. 
Such maxima are recognized as "sideband peaks" in the PE spectrum 
that occur when
\begin{equation}
E_n = n \omega_L + \omega_X -I_p-U_p \label{sideband}
\end{equation}
and they correspond  to the absorption (positive $n$) or emission (negative $n$) of $n$ IR photons 
following the  absorption of one XUV photon. 
In fact, when $N\rightarrow \infty$ the intercycle factor becomes a series of
delta functions, i.e., $ \sum_{n}\delta (E-E_{n})$.

\item[XUV 4] (general case):  
we consider an arbitrary XUV pulse of duration $\tau_X=NT + \Delta$, where $\Delta\leq T$
and delay $t_0 = MT + \delta$, where $\delta\leq T$ (with $N$ and $M$ nonzero integer numbers).
Then, using the previous results we can obtain 
\begin{eqnarray}
T_{if} &=& \int_{MT + \delta }^{MT+\delta + NT+\Delta} \ell(t) e^{iS(t)} dt =e^{i a T M}\int_{\delta}^{\delta + NT+ \Delta}   \cdots
        = e^{i a T M} \left[ \int_0^{NT} + \int_{NT}^{NT+\delta+\Delta} - \int_0^\delta \right]  \label{general}\\
       &=& e^{i a T M} \left[
 I(T) \frac{\sin{(a T N /2)}}{\sin{(a T/2)}}  e^{i a  T (N-1)/2 } +  e^{i a  T N} I(\delta+\Delta) - I(\delta)\right] \quad \textrm{if}\quad \delta + \Delta \leq T \nonumber\\
&=& e^{i a T M} \left[
 I(T) \frac{\sin{(a T (N+1) /2)}}{\sin{(a T/2)}}  e^{i a  T N/2 } +  e^{i a  T (N+1)} I(\delta+\Delta-T) - I(\delta)\right] \quad \textrm{if}\quad \delta + \Delta \geq T .\nonumber
 \label{Tifarbitrary}
\end{eqnarray}
When the XUV covers an integer number of IR cycles ($\Delta =0$), \Eref{general} results in:
\[
T_{if} =  e^{i a T \frac{2M+N-1}{2}} \,\frac{\sin{(a T N /2)}}{\sin{(a T/2)}}\, 
          \left[I(T) + I(\delta)\, 2 i\sin(aT/2) \, e^{\left( i a  T /2\right) } \right],
\] 
where  the intercycle contribution can be factorized as in the previous case. Here we note that at the sideband position $aT=2n\pi$ the second term inside the bracket is zero, then the emission probability is independent of the delay $\delta$ and the PE spectrum becomes  equal to \Eref{intrainter}.
This independence was observed in previous works (see Figure 6 of \cite{Gramajo16} and  5 of \cite{Gramajo17})  as a blurred SCM discontinuity in the PE spectrum as a function of the delay between XUV and IR pulses.
Anyway, in the general case ($\Delta \neq 0$)  we note that the first term inside the brackets  in  \Eref{general}   determines the predominant contribution to the PE spectrum when $N>>1$ 
due to the increase of the  intercycle interference term for $aT=2n\pi$ 
[see discussion below \Eref{sideband}]. In such case, the PE spectrum approximately behaves like \Eref{intrainter}.

\end{description}

Throughout the analysis of the previous four cases, we have covered two LAPE regimes: the streaking (cases 1 and 2) and sideband regimes (cases 3 and 4).
In all these cases,  we found that the temporal integration over the XUV duration  with however delay could be written as function of the magnitude $I(t)$  defined during  the first IR cycle. 
As an example, in Figure \ref{fig0}(b) we show $|I(t)|^2$ for photoionization of Ar($3s$).
From its definition in \Eref{It}, it is clear that it increases from zero at time zero and it depends on the electron energy and the geometrical arrangement between $ \hat{\varepsilon}_X $,  $ \hat{\varepsilon}_L $
 and $ \hat{k}$ (see the arrows in the bottom left corner of the figure).
With this information and employing equations \eref{INT} to \eref{general}, the PE spectra can be constructed
 for several configurations of the XUV+IR fields.
Furthermore, we note that the precedent analysis can also be done not only within the SFA but also within others approaches, such as the Coulomb-Volkov approximation, as long as its dipole elements $\vec{d}[\vec{k}+\vec{A}(t)]$ maintain the temporal $T$-periodicity.


Now, we consider the particular situations in which $\hat{\varepsilon}_L \perp \vec{k}$. 
Because of this, $b=0$ in \Eref{S} and $[S(t)-at]$ has not only $T$- but also $T/2$-periodicity.
If under this circumstance,   the dipole element also satisfies 
$\hat{\varepsilon}_X \cdot\vec{d}[\vec{k}+\vec{A}(t+T/2)] = \pm \hat{\varepsilon}_X \cdot\vec{d}[\vec{k}+\vec{A}(t)] $, \textit{i.e.}, it is symmetric or antisymmetric with respect to the middle  of the IR cycle, 
the integration over all the IR cycle can be written as a sum over the two half cycles.
Then, depending on the symmetry ($+$) or antisymmetry ($-$) of the dipole element with respect to $T/2$ we have 
\begin{eqnarray}
I(T) &=& \int_0^{T/2} \ell(t) e^{iS(t)} dt + \underbrace{\int_{T/2}^T \ell(t) e^{iS(t)} dt}_{\pm e^{iaT/2} I(T/2)} = I(T/2)(1 \pm e^{iaT/2}) \\
|I(T)| &=& |2 ~ I(T/2) ~ \cos{(a T/4)} | \, \, \textrm{ if }+  \, \, \textrm{ (symmetric) } \label{si} \\
|I(T)| &=& |2 ~ I(T/2) ~ \sin{(a T/4)} | \, \, \textrm{ if }- \, \, \textrm{ (antisymmetric) }. \label{asi}
\end{eqnarray}  
The factors $\cos(a T/4)$ or $\sin(a T/4)$  in equations (\ref{si}) and (\ref{asi}) cancel out odd or even sideband peaks in the intercycle contribution, respectively. 
In consequence, the PE spectrum presents structures corresponding to absorption or emission of only even or odd number of IR photons and  the energy difference between two consecutive sideband peaks is $2\omega$ instead of
 $\omega$ as for the general conservation energy rule in \Eref{sideband}.
With this in mind, for antisymmetric dipole elements
the PE spectrum of \Eref{intrainter} becomes
\begin{eqnarray}
|T_{if}|^2&=&  \underbrace{4 ~\underbrace{|I(T/2)|^{2}}_{\textrm{intrahalfcycle}}~ \sin^2(aT/4) }_{\textrm{intracycle}}~
\underbrace{\left[\frac{\sin{( a T N /2)}}{ 2 \sin{(a T/4) \cos(a T /4)}}\right]^2}_{\textrm{intercycle}}
\label{PEasi-1} \\
&=& \underbrace{|I(T/2)|^{2}}_{\textrm{intrahalfcycle}}~\underbrace{\left[\frac{\sin{( a T N /2)}}{ \cos(a T /4) }\right]^2}_{\textrm{interhalfcycle}}, \label{PEasi}
\end{eqnarray}
which  reaches  maxima only for odd $n$ in \Eref{sideband} and it is suppressed at energy values $E_n$ with even $n$.
In particular, the absorption of only one XUV photon alone (in the absence of absorption or
emission of IR photons) is forbidden.
Alternatively, 
the odd sideband orders are canceled whereas the even orders stay on for symmetric dipole elements.
In correspondence with our previous analysis within the SCM (see Eq. (18) in Ref. \cite{Gramajo17}), 
the equations  \eref{PEasi-1} and \eref{PEasi}  
indicate that the PE spectrum can be factorized in two different ways:
(i)  as the product of intra- and intercycle interference factors and
(ii) as the product of intrahalf- and interhalf-cycle interference contributions.
Obviously, the two different factorizations give rise
to the same results.

As an example, we examine again the Ar($3s$) photoionization presented in figure \ref{fig0} (b) and (c). 
The length-gauge dipole element for an hydrogen-like $3s$ state is written in \ref{apen_dipol}.
When both lasers are polarized in the same direction $\hat{z}$ and the electronic emission is perpendicular to them ($\vec{k}=k\hat{x}$), 
the term  $\ell(t)\sim \hat{z} \cdot [k \hat{x} +  A(t)\hat{z}] f_1[k^2+ A^2(t)] = A(t)f_1[k^2+ A^2(t)] $ can be written as the product of  the  vector potential  amplitude $A(t)$ and a function $f_1$ which depends on time through  $A^2(t)$. 
Hence, $\ell(t)$ results antisymmetric respect to the middle of the IR cycle and \Eref{asi} is verified. 

In figure \ref{fig0} (c) we plot the cuts of figure \ref{fig0} (b) at times $t=T/2$ 
and $T$.  
We  observe that, according to \Eref{asi},
 the intracycle contribution $|I(T)|^2$ (orange line) could be written as the product of the intrahalfcycle $|I(T/2)|^2$ (green line) and the  interference factor $\sin^2(aT/4)$ (dashed line) which is null at  even orders of the sideband peaks.

\begin{table}
\caption{Component of dipole element matrix $\hat{\varepsilon}_X \cdot \vec{d}(\vec{v})$ where
 $\vec{v} = \vec{k} + \vec{A}(t)$, and its symmetries respect $T_L/2$, depending on the initial states and the geometrical arrangements. 
}
\footnotesize
\begin{center}
\begin{tabular}{l|llll}
\br
     & F           & P        & OF     & OP    \\
     & $v^2=(k+A)^2$& $v^2= k^2+A^2$ &$v^2= k^2+A^2$& $ v^2=(k+A)^2$        \\
    & $v_z=k+A$   & $v_z= A$&$v_z= k$& $ v_z=0$     \\  
\mr
$\vec{d}_{3s}\cdot \hat{z}=v_z f_1(v^2)$& $=(k+A)f_1((k+A)^2)$ & $=A f_1(k^2+A^2)$ & $=k f_1(k^2+A^2)$& $ =0 $\\
$3s$ & no symmetry               & antisymmetric      &  symmetric         &                      \\
    &   all SB                  & odd SB             & even SB            & not  SB            \\    
\hline
$\vec{d}_{3p0}\cdot \hat{z} =f_2(v^2)+ f_3(v^2)v_z^2$    & $\sim (k+A)^2$& $\sim A^2$ &$\sim A^2$  & $=f_2((k+A)^2)$\\
& no symmetry          & symmetric        &  symmetric  & no symmetry      \\
$3p0$   &   all SB             & even SB          & even SB     & all SB           \\ 
\hline
$\vec{d}_{3p1}\cdot\hat{z} = v_z(v_x+iv_y)f_4(v^2)$     & $=0 $& $= A k f_4(k^2+A^2)$& $ =  A k f_4(k^2+A^2)$& $=0 $ \\
&           & antisymmetric        & antisymmetric        &                 \\
$3p1$   & not SB    & odd SB               & odd SB               & not SB       \\ 
\br
\end{tabular}
\end{center}
\label{t1}
\end{table}
\normalsize

\section{LAPE of argon atoms}

In this section we want to explore in what situations some  sideband orders are canceled in view of the preceding discussion.
For that, we  consider the argon photoionization from the shell 3,  and analyze the symmetries of the states $3s$, $3p_0$ and $3p_1$
 for different geometrical arrangements between the momentum direction  $\hat{k}$ and the polarization vectors of both linearly polarized fields. 
We fix  $\hat{\varepsilon}_X$  along the $\hat{z}$ axis, whereas the IR polarization vector $\hat{\varepsilon}_L$  can be parallel or orthogonal to it.
For the sake of simplicity, we restrict our analysis to the case where the XUV pulse duration is a multiple of the IR optical cycle, i.e., $\tau_{X} = N T$. 
Then, the PE momentum distribution is given by \Eref{intrainter}, and
we examine the emission parallel and perpendicular  to the XUV polarization direction.
Different combinations of these geometrical arrangements lead to  four study cases, namely:

\begin{figure}[h!]
\centering
\includegraphics[angle = 0, width=0.90 \textwidth]{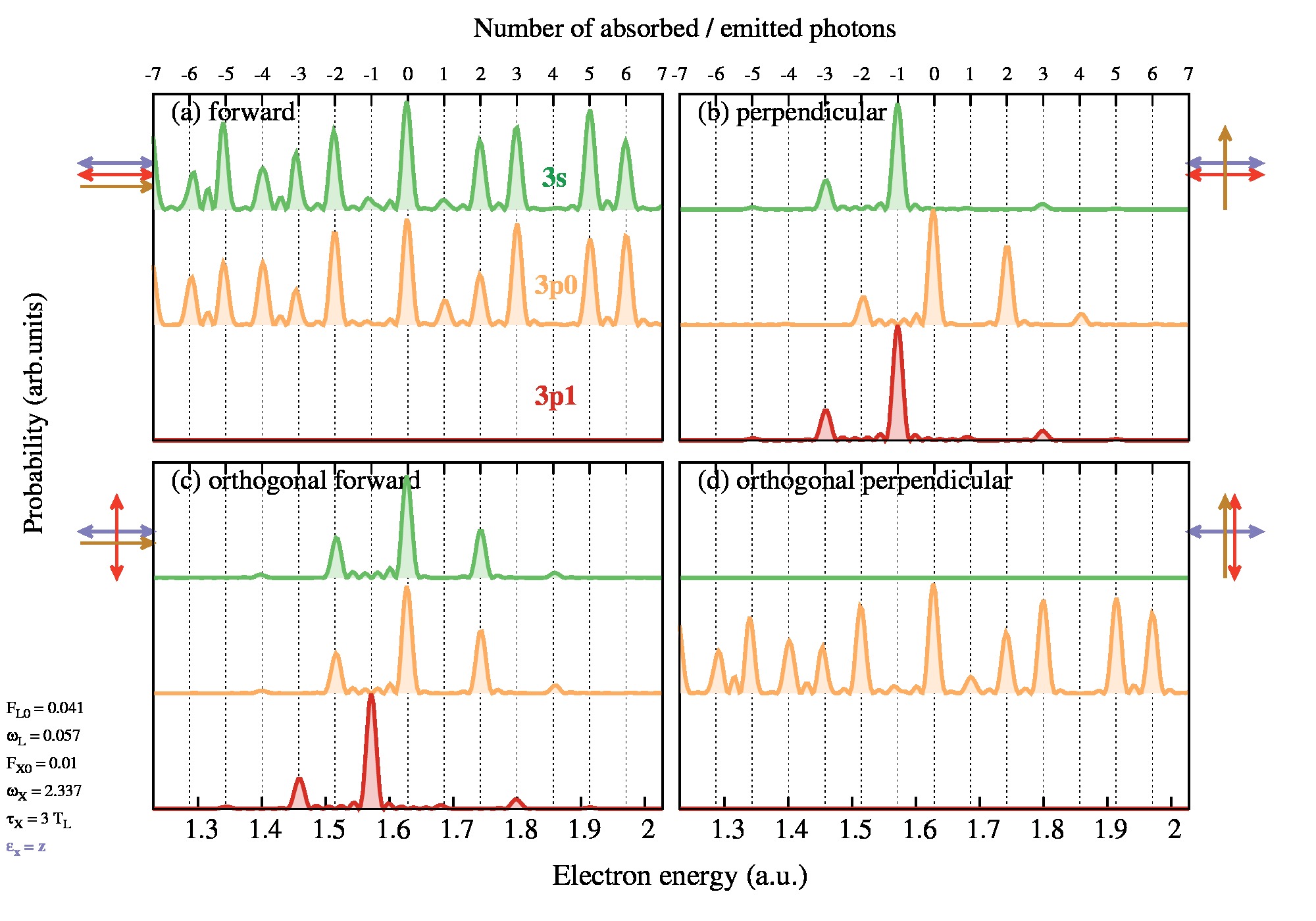}
\caption{PE spectra in arbitrary units for LAPE of Argon from shell 3. 
The arrows represent the geometrical arrangement between $\hat{\varepsilon}_X$ (blue), $\hat{\varepsilon}_L$ (red) and $\hat{k}$ (orange). The laser parameters are the same as in Fig. 1
}
\label{fig1}
\end{figure}

\begin{description}
\item[Forward (F):] Both polarization vectors and electronic emission direction are  parallel to the $\hat{z}$ axis, \textit{i.e.}, $\hat{\varepsilon}_L= \hat{\varepsilon}_X = \hat{k} =  \hat{z}$. 
\item[Perpendicular (P):] Both polarization vector are parallel and the electron emission direction is perpendicular to them, \textit{ i.e.}, $\hat{\varepsilon}_L= \hat{\varepsilon}_X  = \hat{z}$ and $\hat{k} =  \hat{x}$.
\item[Orthogonal Forward (OF):] The laser polarizations are orthogonal and the electron emission is parallel to the XUV one, \textit{ i.e.}, $\hat{\varepsilon}_X= \hat{k} = \hat{z}$ and $\hat{\varepsilon}_L =  \hat{x}$.
\item[Orthogonal Perpendicular (OP):] The laser polarizations are orthogonal and the electron emission is parallel to the IR one, \textit{i.e.},
$\hat{\varepsilon}_X= \hat{z}$ and $\hat{\varepsilon}_L =\hat{k}=  \hat{x}$.
\end{description}

The hydrogen-like dipole matrix elements for the subshells  $3s$, $3p0$ and $3p1$ are described in  \ref{apen_dipol}. 
In table \ref{t1} we present the analysis of the expected sideband (SB) orders 
depending on the geometrical arrangement (columns) and the dipole element $\hat{z}$-component (rows) evaluated in $\vec{v}=\vec{k}+\vec{A}(t)$.
For each geometrical configuration the magnitudes  $v^2$ and $v_z$ are shown in the first row. 
In F and OP configurations, the temporal dependence of $v^2= [k+A(t)]^2$  does not have any particular  symmetry  for what we would expect all sideband orders according to \Eref{intrainter}. 
In particular, (F $3p1$), (OP $3s$) and (OP $3p1$) have null dipole elements.
Instead, the  P and OF configurations (center columns) verify the conditions $b=0$ (since $\hat{\epsilon}_L \perp \vec{k}$)  and there is a defined symmetry 
 of the dipole element (symmetric or antisymmetric) with
respect to $T/2$.
Thus,  even or odd SB orders are expected.

In order to corroborate the theoretical predictions of table 1, we also present 
in Fig. \ref{fig1}
the SFA calculation of Eq. (\ref{Tifg}). 
The four geometrical configurations P, F, OF and OP are considered in figures \ref{fig1} (a) to (d), respectively,  and for the initial states   $3s$ (in green) $3p_0$ (yellow) and $3p_1$ (red). 
The numbers $n$ of absorbed or emitted photons are indicated as vertical dashed lines in the figure. 
Since the difference of ionization potential $I_p$
for the different argon initial states considered
 (see \ref{apen_dipol}), the $3s$-PE spectrum is shifted $0.48$ a.u. to high energies in order to match the same sideband order for all states.
In the figure we observe that the sideband peaks have different heights, because they 
depend on the modulation of the intracycle factor  \cite{Gramajo16, Gramajo17}. 
Furthermore, in agreement with the analysis of table \ref{t1} we note  that:
(i)  in P and OF configurations only odd or even sideband are present, whereas
(ii) the F and OP configurations exhibit  all SB orders, except when the emission is forbidden (F $3p1$, OP $3s$ and $3p1$) or when the intracycle modulation suppress some  particular SB orders ($n=-1$ and $4$ for example).
Besides, (iii) the energy range of the emitted electrons is much smaller in P and OF arrangements than in the F and OP cases,
strongly depending  on the IR intensity \cite{Gramajo18}.
In addition, we want to emphasize that TDSE calculations for Argon (P $3s$) 
corroborate the selection of only odd sideband peaks allowed in the PE spectrum, as it can be observed from  Fig.  1 of Ref. \cite{posterIcpeac19}.


\section{\label{conc}Conclusions}


We have studied the electron emission produced by an XUV pulse assisted by an IR laser field
emphasizing on  the analytic properties deduced from 
the SFA transition matrix element. 
We have shown  that in several XUV+IR configurations,
the PE spectrum can be described as a function of the time integral $I(t)$ during the first IR optical cycle
not only in the sideband but also in the streaking regimen.
In particular, we have shown that intra-, inter-, intrahalf- and interhalfcycle interferences are a consequence of the periodicity and symmetry of the transition matrix element. 
For the case of   photoionization of argon from the third shell,
we have analyzed the symmetries of the states $s$, $p_0$ and $p_1$ in four different geometrical arrangements and have corroborated the corresponding selection rules  that determine the presence of all, none, odd or even sideband orders in the PE spectra.

\appendix
\section{Dipole elements}\label{apen_dipol}

 The dipole element is defined by 
$\vec{d}_i(\vec{v}) = (2\pi)^{-3/2}  \int d \vec{r} \, \exp[-i \vec{v}\cdot \vec{r}]\,  \vec{r} \, \phi_i(\vec{r})$.
Depending on the initial hydrogen-like state $\phi_i$ we have 
that the $\hat{z}$-component of the dipole elements  are
%
%
\begin{displaymath}
\begin{array}{lll}
\vec{d}_{3s} \cdot\hat{z} =            v_z          f_1(v^2)  
&\quad& f_1(v^2) =       -2i \, (2\alpha)^{5/2}\, (3 v^4 +11\alpha^4 - 18 \alpha^2 v^2 )
                                                                  / \pi(v^2+\alpha^2)^5  
\\
\vec{d}_{3p0}\cdot\hat{z} = f_2(v^2) + v_z^2        f_3(v^2) 
&\quad& f_2(v^2) = \sqrt{3} \, 2^{4}  \alpha^{7/2}\, (v^4-\alpha^4 )    / \pi  (v^2+\alpha^2)^5 
\\
&& f_3(v^2) =-\sqrt{3} \, 2^{5}  \alpha^{7/2}\, (3 v^2 -5 \alpha^2)/ \pi  (v^2+\alpha^2)^5
\\
\vec{d}_{3p1}\cdot\hat{z} =            v_z(v_x+iv_y)f_4(v^2) 
&&f_4(v^2) = \sqrt{3} \, 2^{9/2}\alpha^{7/2}\, (3 v^2-5 \alpha^2) / \pi  (v^2+\alpha^2)^5
\end{array}
\end{displaymath}
where $\alpha = \sqrt{2I_p}$ and we have introduced the functions 
$f_j$ (for $j=1$ to 4) to indicate explicitly the dependence on  $v^2$. 
In order to consider the photoionization of argon, we have set  the ionization potential  $I_p =15.78$ eV for the $3p$ subshells and $I_p =28.84$ eV for the $3s$ state.


\section*{References}
\bibliography{iopart-num}

\end{document}